\documentclass[twocolumn,amssymb, nobibnotes, showpacs, superscriptaddress, aps, prd]{revtex4}
\usepackage{amsmath}
\usepackage{epsfig}

\usepackage{graphicx}% Include figure files
\usepackage{dcolumn}% Align table columns on decimal point
\usepackage{bm}% bold math
\usepackage[bookmarks=true,
   colorlinks=true,
   linkcolor=blue,
   urlcolor=blue,
   citecolor=blue,
   bookmarks=true,
   hyperindex=true
]{hyperref}

\begin{document}

\title{Rabi oscillation induced $\pi$-phase flip in an unbalanced Ramsey atom interferometer}

\author{R. B. Li}
\email[]{rbli@wipm.ac.cn}
\affiliation{State Key Laboratory of Magnetic Resonance and Atomic and Molecular Physics,
Wuhan Institute of Physics and Mathematics, Chinese Academy of Sciences, Wuhan
430071, China}
\affiliation{Center for Cold Atom Physics, Chinese Academy of Sciences, Wuhan 430071, China}
\author{Z. W. Yao}
\affiliation{State Key Laboratory of Magnetic Resonance and Atomic and Molecular Physics, Wuhan Institute of Physics and Mathematics, Chinese Academy of Sciences, Wuhan 430071, China}
\affiliation{Center for Cold Atom Physics, Chinese Academy of Sciences, Wuhan 430071, China}
\author{K. Wang}
\affiliation{State Key Laboratory of Magnetic Resonance and Atomic and Molecular Physics, Wuhan Institute of Physics and Mathematics, Chinese Academy of Sciences, Wuhan 430071, China}
\affiliation{Center for Cold Atom Physics, Chinese Academy of Sciences, Wuhan 430071, China}
\affiliation{University of Chinese Academy of Sciences, Beijing 100049, China}
\author{S. B Lu}
\affiliation{State Key Laboratory of Magnetic Resonance and Atomic and Molecular Physics, Wuhan Institute of Physics and Mathematics, Chinese Academy of Sciences, Wuhan 430071, China}
\affiliation{Center for Cold Atom Physics, Chinese Academy of Sciences, Wuhan 430071, China}
\affiliation{University of Chinese Academy of Sciences, Beijing 100049, China}
\author{L. Cao}
\affiliation{State Key Laboratory of Magnetic Resonance and Atomic and Molecular Physics, Wuhan Institute of Physics and Mathematics, Chinese Academy of Sciences, Wuhan 430071, China}
\affiliation{Center for Cold Atom Physics, Chinese Academy of Sciences, Wuhan 430071, China}
\affiliation{University of Chinese Academy of Sciences, Beijing 100049, China}
\author{J. Wang}
\email[]{wangjin@wipm.ac.cn}
\affiliation{State Key Laboratory of Magnetic Resonance and Atomic and Molecular Physics,
Wuhan Institute of Physics and Mathematics, Chinese Academy of Sciences, Wuhan
430071, China}
\affiliation{Center for Cold Atom Physics, Chinese Academy of Sciences, Wuhan 430071, China}

\author{M. S. Zhan}
\email[]{mszhan@wipm.ac.cn}
\affiliation{State Key Laboratory of Magnetic Resonance and Atomic and Molecular Physics,
Wuhan Institute of Physics and Mathematics, Chinese Academy of Sciences, Wuhan
430071, China}
\affiliation{Center for Cold Atom Physics, Chinese Academy of Sciences, Wuhan 430071, China}

\date{\today}

\begin{abstract}
We present an observation of zero to $\pi$ phase flips induced by Rabi oscillation in an unbalanced Ramsey atom interferometer. The phase shift and visibility are experimentally investigated by modulating either the polarization or duration of Raman lasers, and they are well explained by a theoretical model. In an atom interferometer, the $\pi$-phase flips are caused not only by the sign of Rabi frequency but also by the Rabi oscillation. Considering the $\pi$-phase flips, we propose the composite-light-pulse sequences for realizing the large-momentum-transfer beam splitter and mirror, which have the high immunity to the external phase noise in building the cold atom interferometer.
\end{abstract}

%\keywords{}

\maketitle

Since the light-pulse atom interferometer was demonstrated \cite{Kasevich1991a}, it has been applied in many scientific and technical fields, such as measurements of the gravitational constant \cite{Fixler2007a,Rosi2014a} and the fine structure constant \cite{Weiss1994a,Bouchendira2011a}, test of the weak equivalence principle \cite{Schlippert2014a,Tarallo2014a,Zhou2015a}, and inertial instruments of atomic gyroscopes \cite{Canuel2006a,Stockton2011a,Dutta2016a} and atomic gravimeters \cite{Peters1999a,Hu2013a}. In addition, atom interferometers used in test of the general relativity \cite{Dimopoulos2007a,Dimopoulos2008a,Aguilera2014a} and detection of the gravitational wave \cite{Dimopoulos2008b,Hogan2011a,Graham2013a,Chaibi2016a}, were even proposed. The sensitivity of atom interferometer can be improved by enlarging its interferometric area. To make a large area, many techniques including the multi-light-pulse sequence \cite{McGuirk2000a}, sequential multiphoton Bragg diffraction \cite{Muller2008a,Chiow2011a,Kovachy2015a} and frequency-swept Raman adiabatic passage \cite{Kotru2015a} were applied. On the other hand, the sensitivity can be improved by suppressing the phase noise during the atomic free evolution. For this purpose, the double diffraction technique was developed \cite{Leveque2009a,Giese2013a}.

The beam splitter and mirror are the central elements of atom interferometer, and their phase noise limits the quantum behavior determined sensitivity and stability. To further improve the sensitivity of atom interferometer, it is essential to synchronously enlarge the interferometric area and suppress the phase noise. Thus the composite-light-pulse (CLP) sequences were used to create the large-momentum-transfer (LMT) beam splitter and mirror in the atom interferometer \cite{Berg2015a}. In the CLP sequences based atom interferometer, the external technical noise during the intervals of the atomic free evolution, was greatly suppressed due to the atoms stayed in the same states. However, the phase noise in the CLP sequences can not be suppressed during the atomic wave-packet manipulations. Therefore, it is important to simultaneously suppress the phase noise in the CLP sequences and during the atomic free evolution. Recently, we found a $\pi$-phase flip when the polarization, intensity or duration of Raman lasers were modulated in the atom interferometer, which is useful for building the CLP sequences.

In this paper, we design and implement an experiment to quantitatively investigate the phase shift induced by the Rabi oscillation in the atom interferometer. The zero to $\pi$ phase flips are observed in an unbalanced Ramsey atom interferometer by modulating the polarization and duration of Raman lasers, and they are well explained by the sign of Rabi frequency and the Rabi oscillation. Considering the $\pi$-phase flips between two consecutive Raman transitions, the new CLP sequences are proposed for building the LMT beam splitter and mirror. The phase noise is discussed, and it can be cancelled in the CLP sequences and during the atomic free evolution. These CLP sequences can be used to build the low-phase-noise large-area atom interferometer, and it is important to improve the sensitivity and stability of atom interferometer.

\begin{figure}[htp]
	\centering
		\includegraphics[width=0.38\textwidth]{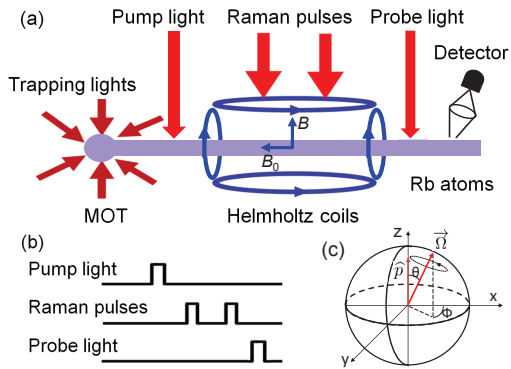}
	\caption{(Color online) Experiment setup (a), timing sequences (b) and Bloch sphere depiction (c). Cold atoms are propagating from the MOT to the probe region. The atoms are manipulated by two pairs of Raman lasers with a time interval of $T$. Ramsey fringes are observed by measuring the population using the laser-induced fluorescence signal.}%
	\label{fig-1}%
\end{figure}

The experiment is performed in the unbalanced Ramsey interferometer where the atoms are coherently manipulated by two asymmetrical Raman pulses \cite{Note}. The experimental arrangement (top view) is shown in Fig.\textit{\ref{fig-1}} (a). The atoms ($^{85}$Rb) are trapped in a magneto-optical trap (MOT). Then they are launched along a parabolic trajectory, with an initial velocity of 2.5 m/s and an angle of 14.0 $^{\circ}$ with respect to the gravity. The polarization gradient cooling is applied when the atoms are accelerating in the MOT. Thus, an atom fountain is built with the low temperature ($<$5 $\mu$K). When cold atoms are propagating from the MOT to probe regions, they are pumped to the state $|5S_{1/2},F=2\rangle$. Two pairs of Raman lasers, with a time interval of $T=50$ ms, are co-propagating along the magnetic field $B$ (100 mG). The magnetic field $B_{0}$ is adjusted to cancel the residual magnetic field \cite{Li2008a}. The Raman lasers with frequency difference of 3.04 GHz are generated from the $\pm$1 order diffracted lights of an acousto-optic modulator (1.52 GHz) \cite{Wang2016a}. The Raman lasers couple two ground states $|F=2,m_{F}=0\rangle$ and $|F=3,m_{F}=0\rangle$ with the excited states $|5P_{3/2}\rangle$, where the one-photon detuning is 1.5 GHz. When the atoms are operated by the Raman lasers, Raman transitions are observed and Raman pulses are defined by the Rabi frequency and duration of Raman lasers. When the atoms in the initial state $|F=2,m_{F}=0\rangle$ are manipulated by two symmetrical or asymmetrical Raman pulses, the balanced and unbalanced Ramsey fringes are observed by measuring the population in the state $|F=3,m_{F}=0\rangle$ using the laser-induced fluorescence signal. 

\begin{figure}[htp]
	\centering
		\includegraphics[width=0.36\textwidth]{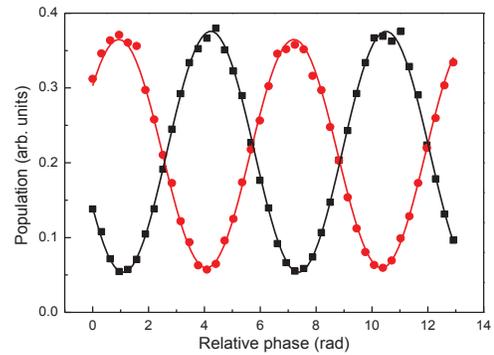}
	\caption{(Color online) Observed Ramsey fringes with two Raman pulses of $\left(\sigma^{+},\sigma^{+}\right)$ and $\left(\sigma^{+},\sigma^{+}\right)$ (black squares), while $\left(\sigma^{+},\sigma^{+}\right)$ and $\left(\sigma^{-},\sigma^{-}\right)$ (red circles). There is a $\pi$-phase shift when the polarization is rotated from $\left(\sigma^{+},\sigma^{+}\right)$ to $\left(\sigma^{-},\sigma^{-}\right)$.}%
	\label{fig-2}%
\end{figure}

\begin{figure}[htp]
	\centering
		\includegraphics[width=0.39\textwidth]{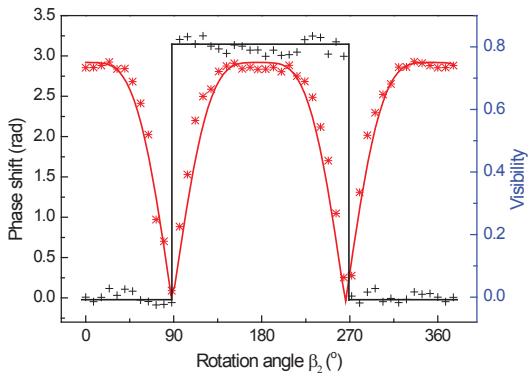}
	\caption{(Color online) Phase shift (black crosses) and visibility (red stars) as a function of the polarization rotation angle $\beta_{2}$. They are fitted by the theoretical model (solid curves).}%
	\label{fig-3}%
\end{figure}

With the timing sequences in Fig.\textit{\ref{fig-1}} (b), Ramsey fringes are obtained by scanning the phase difference of Raman lasers, as shown in Fig.\textit{\ref{fig-2}}. A typical balanced Ramsey fringe is observed with two symmetrical $\pi/2$ pulses of $\left(\sigma^{+},\sigma^{+}\right)$ and $\left(\sigma^{+},\sigma^{+}\right)$ (black squares), and an unbalanced Ramsey fringe is also observed with two asymmetrical $\pm\pi/2$ pulses of $\left(\sigma^{+},\sigma^{+}\right)$ and $\left(\sigma^{-},\sigma^{-}\right)$ (red circles). There is a $\pi$-phase shift when the polarization of the second pulse is rotated from $\left(\sigma^{+},\sigma^{+}\right)$ to $\left(\sigma^{-},\sigma^{-}\right)$. In the same way, the $\pi$-phase shift is also observed when the polarization of the first pulse is rotated from $\left(\sigma^{+},\sigma^{+}\right)$ to $\left(\sigma^{-},\sigma^{-}\right)$ while the polarization of the second pulse is maintained in $\left(\sigma^{+},\sigma^{+}\right)$.
The phase shift and visibility are quantitatively investigated when the polarization is continuously rotated in one cycle. Here, the first $\pi/2$ pulse stays in $\left(\sigma^{+},\sigma^{+}\right)$, and the polarization of the second $\pi/2$ pulse is rotated from $\beta_{2}=0$ to $360^{\circ}$ by rotating the polarization axis of the $\lambda$/4 wave plate from $\beta_{2}^{'}=0$ to $180^{\circ}$ \cite{Explain1}. Similar to Fig.\textit{\ref{fig-2}}, the Ramsey fringes are obtained and fitted by sine wave functions. Fig.\textit{\ref{fig-3}} shows the phase shift (black crosses) and the visibility (red stars) when the polarization of the second Raman pulse is rotated from $\beta_{2}=0$ to 360$^{\circ}$. In Fig.\textit{\ref{fig-3}} (black crosses), the $\pi$-phase flip is observed by rotating the polarization, which is similar to the results obtained by rotating the magnetic field \cite{Usami2007a,Takahashi2009a}. In their experiments, the observed $\pi$-flipped phase was interpreted as a geometric phase \cite{Usami2007a}, caused by a parity-dependent phase factor of $(-1)^F$ for $m=0$ spin states \cite{Robbins1994a}, and it was also explained by an opposite sign between two transition amplitudes \cite{Takahashi2009a}.

The above results can be well explained by the theoretical equations of the unbalanced Ramsey interferometer. We consider a pair of Raman lasers couple two ground states $|a\rangle$ and $|b\rangle$ with an excited state $|i\rangle$. The first $\pi/2$ pulse takes the initial state $|a\rangle$ into a superposition state ($|a\rangle$+$|b\rangle$)/$\sqrt{2}$. After a free evolution time of $T$, the second arbitrary pulse then recombines the atomic wave packets. Finally, the probability amplitude of the state $|b\rangle$ is obtained by \cite{Ramsey1950a,Borde1984a}
\begin{equation}
b=\frac{1}{\sqrt{2}}[\cos\frac{\Omega_{2}\tau_{2}}{2}e^{-i(\omega_{2}T-\phi_{1})}+\sin\frac{\Omega_{2}\tau_{2}}{2}e^{-i(\omega_{1}T-\phi_{2})}]
\end{equation}
where, $\phi_{j}$ is the phase of the $j$th Raman pulse, $\Omega_{j}$ and $\tau_{j}$ are the Rabi frequency and the pulse duration of the $j$th Raman interactions, and $\omega_{1}$ and $\omega_{2}$ are the perturbed frequencies of two ground states. The population probability of the state $|b\rangle$ is written as
\begin{equation}\label{eq2}
	bb^{*}=\frac{1}{2}\{1+\sin(\Omega_{2}\tau_{2})\cos[(\omega_{2}-\omega_{1})T+\phi_{2}-\phi_{1}]\}
\end{equation}
where, $|\sin(\Omega_{2}\tau_{2})|$ is the visibility. Its behavior can be visually depicted on a Bloch sphere \cite{Bateman2007a,Kotru2014a} as shown in Fig.\textit{\ref{fig-1}} (c). For an given angle $\beta_{2}$, the input polarization has to be transformed into a coordinate system along the magnetic field $B$ using a rotation matrix $D_{QQ^{'}}^{(K)}$, where $K$ is multiplicity and $Q$ is $m_{F}-m_{F^{'}}$ \cite{Wynands1998a}. For $K=1$ and $Q=0$, the effective Rabi frequency is given by \cite{Li2008a}
\begin{equation}
\Omega_{2}(\beta_{2})=\frac{|e|^{2}\varepsilon\xi}{4\hbar^{2}\Delta}\underset{i}{%
%TCIMACRO{\dsum }%
%BeginExpansion
{\displaystyle\sum}
%EndExpansion
}\left\langle b\right\vert r_{q}\left\vert i\right\rangle \left\langle
i\right\vert r_{p}\left\vert a\right\rangle \cos\beta_{2}\label{eq3}%
\end{equation}
where, $\Delta$ is the one-photon detuning, $\left\langle b\right\vert r_{q}\left\vert i\right\rangle \left\langle
i\right\vert r_{p}\left\vert a\right\rangle$ is the reduced matrix element of two-photon Raman transitions, and $\varepsilon$, $\xi$ are the amplitudes of two Raman laser fields, respectively.
When the polarization or magnetic field is rotated as $\beta_{2}$, the sign of $\Omega_{2}(\beta_{2})$ depends on the sign of $\cos\beta_{2}$ term in Eq.(\ref{eq3}). The $\pi$-phase flip is caused by the positive or negative sign of $\sin(\Omega_{2}\tau_{2})$ term in Eq.(\ref{eq2}). The $\pi$-phase flip observed with the polarization rotated in Figs.\textit{\ref{fig-2}} and \textit{\ref{fig-3}} (black crosses) and the magnetic field rotated in Refs.\cite{Usami2007a,Takahashi2009a} can be better explained by the opposite sign of Rabi frequency rather than by the geometric phase. The $\pi$-phase flip caused by the negative sign of transition amplitude is confirmed by modulating the polarization of Raman lasers. Due to two $\pi/2$ Raman pulses initially applied for $\beta_{2}=0$, the visibility can be rewritten as a function of $\left|sin\left(\frac{\pi}{2}cos\beta_{2}\right)\right|$, according to Eqs.(\ref{eq2}) and (\ref{eq3}), when $\beta_{2}$ is rotated. The experimental result (red stars) is well matched with the theoretical model (red solid curve) as shown in Fig.\textit{\ref{fig-3}}.

\begin{figure}
	\centering
		\includegraphics[width=0.38\textwidth]{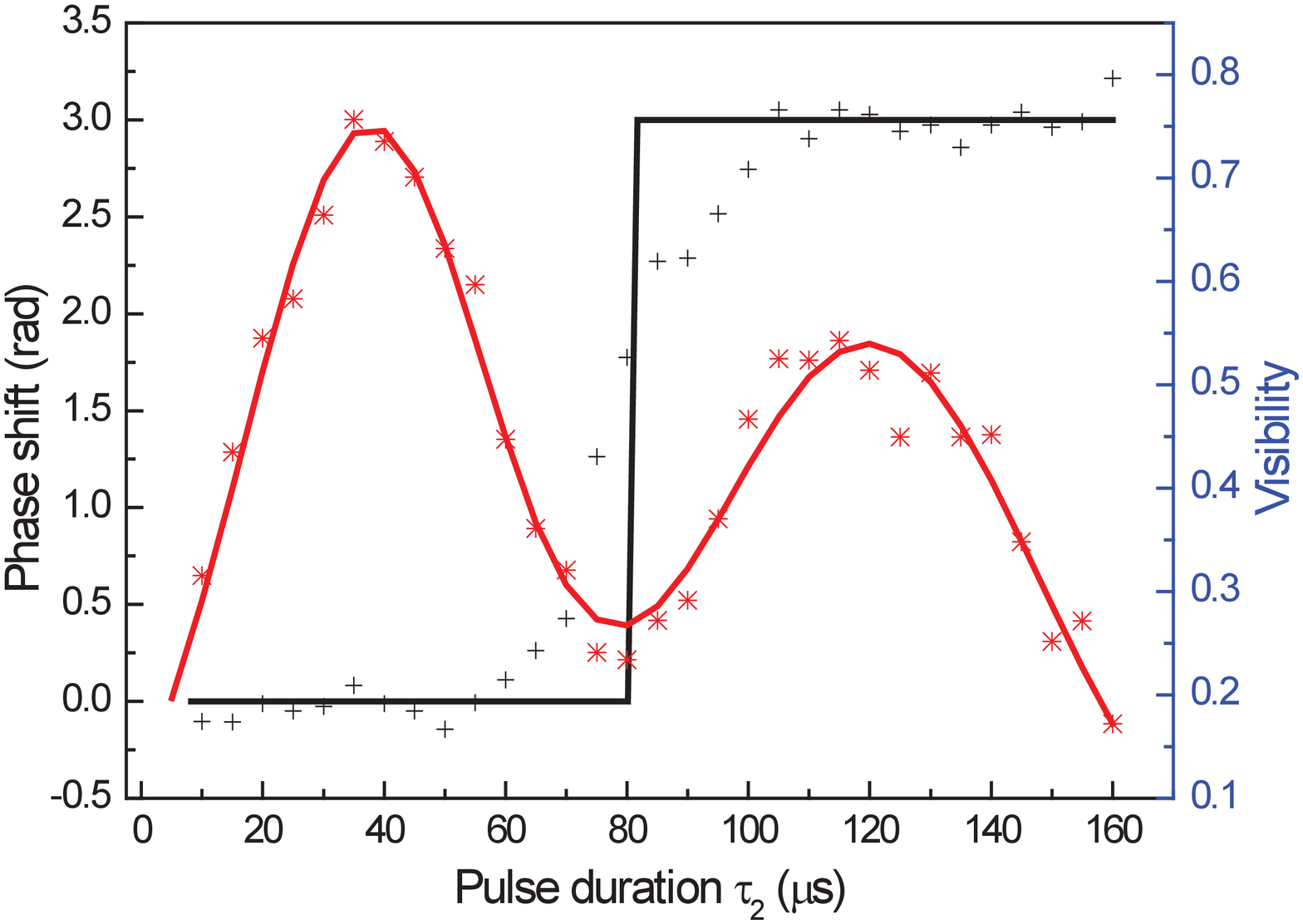}
	\caption{(Color online) Phase shift (black crosses) and visibility (red stars) as a function of the pulse duration $\tau_{2}$. There is a $\pi$-phase flip when the pulse duration is 80 $\mu$s.}%
	\label{fig-4}%
\end{figure}

The phase shift and visibility are investigated by modulating the pulse duration ($\tau_{2}$). In the experiment, two pulses of $\left(\sigma^{+},\sigma^{+}\right)$ and $\left(\sigma^{+},\sigma^{+}\right)$ with the same Rabi frequencies ($\Omega_{1}=\Omega_{2}$) are co-propagating along the magnetic field $B$. After the atoms are prepared to the coherent superposition state by the first pulse ($\Omega_{1}\tau_{1}=\pi/2$), they are manipulated by the second pulse whose duration is increased from $\tau_{2}=0$ to 160 $\mu$s. For the first time, the Rabi oscillation induced $\pi$-phase flip is experimentally observed by modulating the duration as shown in Fig.\textit{\ref{fig-4}} (black crosses). When the duration is increased, the Raman pulse is equivalent to a combination of two Raman pulses. For example, the second pulse (for $\Omega_{2}\tau_{2}=3\pi/2$) is equivalent to the combination of $\pi$ and $\pi/2$ pulses. The former exchanges the population between two coherent states, and the latter recombines the atomic wave packets. Therefore, this $\pi$-phase flip can be well explained by the population exchanges between two coherent states. For the different durations, the visibility is shown in Fig.\textit{\ref{fig-4}} (red stars). From Eq.(\ref{eq2}), the visibility should be maximum for $\Omega_{2}\tau_{2}=\pm M\pi/2$ ($M$ is an odd number), and Ramsey fringes are vanished for $\Omega_{2}\tau_{2}=\pm N\pi/2$ ($N$ is an even number). The maximum visibility is observed for $\tau_{2}=40$ $\mu$s ($\Omega_{2}\tau_{2}=\pi/2$). The Ramsey fringes are nearly disappeared for $\tau_{2}=80$ $\mu$s ($\Omega_{2}\tau_{2}=\pi$) and $\tau_{2}=160$ $\mu$s ($\Omega_{2}\tau_{2}=2\pi$) because the atomic wave packets are not recombined. Due to the atomic velocity distribution, the decoherence is caused by the spatial intensity variation and wave front distortion of Raman lasers. The visibility for $\tau_{2}=120$ $\mu$s ($\Omega_{2}\tau_{2}=3\pi/2$) is less than that for $\tau_{2}=40$ $\mu$s. The visibility is decreased for other durations because of the partial recombination of wave packets, which is consistent with the function of $|\sin(\Omega_{2}\tau_{2})|$.

In above experiments, the observed $\pi$-phase flips arise from the positive or negative sign of $\sin(\Omega_{2}\tau_{2})$ term in Eq.(\ref{eq2}). This behavior can be visualized on the Bloch sphere as shown in Fig.\textit{\ref{fig-1}} (c). The drive field ($\overrightarrow{\Omega}$) is rotated at a polar angle ($\theta=\Omega\tau$) when the Raman pulse is applied. The first pulse ($\theta_{1}=\pi/2$) drags the Bloch vector $\widehat{p}$ at the positive $z$-axis into the $x$-$y$ plane. After the integration time $T$, $\widehat{p}$ is then rotated at $\theta_{2}=\Omega_{2}\tau_{2}$ on the Bloch sphere when the second pulse is applied. The azimuthal angle ($\Phi=(\omega_{2}-\omega_{1})T+\phi_{2}-\phi_{1}$) is rotated around $z$-axis by scanning the phase difference of the Raman lasers and it is same for each Ramsey fringe due to the cancellation of the ac Stark shift \cite{Li2009a} and the synchronous rotation of polarizations \cite{Explain2}. The projection of $\widehat{p}$ onto the $z$-axis, $\sin\theta_{2}$, gives the population difference, which can depict the phase and visibility of the fringe. When $\beta_{2}$ is rotated, an elliptical polarization is equivalent to a combination of linear and circular polarizations. Because the two-photon transition probability induced by the linear polarization can be neglected for the large one-photon detuning, the second pulse is equivalent to  $\left(\sigma^{+},\sigma^{+}\right)$ or $\left(\sigma^{-},\sigma^{-}\right)$. The signs between two Rabi frequencies are opposite for $\left(\sigma^{+},\sigma^{+}\right)$ and $\left(\sigma^{-},\sigma^{-}\right)$. For the positive or negative Rabi frequency, $\widehat{p}$ is projected on the negative or positive $z$-axis, which explains the $\pi$-phase flips in Fig.\textit{\ref{fig-3}} (black crosses). When $\tau_{2}$ is increased, $\widehat{p}$ is rotated from the $x$-$y$ plane to the negative $z$-axis. $\widehat{p}$ is projected on the negative $z$-axis when $\theta_{2}$ is changed from zero to $\pi$, while it is projected on the positive $z$-axis when $\theta_{2}$ is changed from $\pi$ to $2\pi$. Thus the $\pi$-phase flip is observed as shown Fig.\textit{\ref{fig-4}} (black crosses). The visibility is maximum when $\widehat{p}$ is parallel to $z$-axis, and the fringe is disappeared when $\widehat{p}$ is vertical to $z$-axis.

\begin{figure}[htp]
	\centering
		\includegraphics[width=0.36\textwidth]{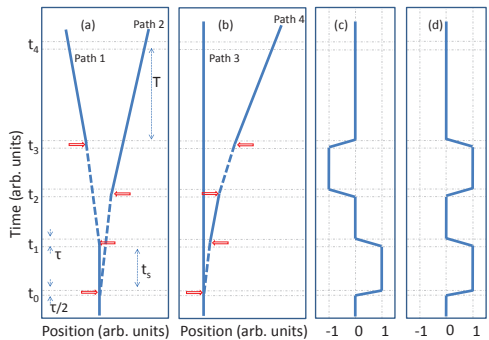}
	\caption{(Color online) A space-time diagram of the LMT beam splitter for symmetric (a) and asymmetric (b) CLP sequences as described in Refs.\cite{Berg2015a,Graham2013a}, and the corresponding sensitivity functions with (c) and without (d) considering the Rabi oscillation and sign induced $\pi$-phase flips. The sensitivity functions are calculated by the theoretical model in Ref.\cite{Cheinet2005a}}%
	\label{fig-5}%
\end{figure}

When the CLP sequences are applied to build the LMT beam splitter and mirror in atom interferometers, the $\pi$-phase flip modulated by the sign of Rabi frequency or the Rabi oscillation is useful for cancelling the slow varying phase noise. We discuss the phase noise in the CLP sequences composed of several rapid-succession Raman pulses, where a pair of Raman lasers are counter-propagating for the larger momentum transfer. Fig.\textit{\ref{fig-5}} (a) shows a LMT beam splitter of the symmetric CLP sequence, but four pulses are applied instead of two pulses in Ref.\cite{Berg2015a}. Fig.\textit{\ref{fig-5}} (b) shows a LMT beam splitter of the asymmetric CLP sequence as proposed in Ref.\cite{Graham2013a}. At the time $t_{0}$, the atoms in the initial state $|a\rangle$ are split into the states $|a\rangle$ and $|b\rangle$ by the first $\pi/2$ pulse. From $t_{1}$ to $t_{3}$, three $\pi$ pulses alternately manipulate the states of atoms in the path 1 or 2 as shown in Fig.\textit{\ref{fig-5}} (a), while they only manipulate the states of atoms in the path 4 as shown in Fig.\textit{\ref{fig-5}} (b). Due to the wave vectors reversed (red double arrows), the 8$\hbar k$ LMT beam splitters can be built with the symmetric and asymmetric CLP sequences.

When the pulse duration ($\tau$) is $10$ $\mu$s and the dark time ($t_{s}$) is $20$ $\mu$s \cite{Berg2015a}, the LMT beam splitters of the symmetric and asymmetric CLP sequences can be realized in $100$ $\mu$s (from $t_{0}$ to $t_{3}$). The sensitivity functions are calculated by the theoretical model \cite{Cheinet2005a}. Fig.\textit{\ref{fig-5}} (c) plots the sensitivity function of the symmetric CLP sequence in Fig.\textit{\ref{fig-5}} (a). Fig.\textit{\ref{fig-5}} (d) is the sensitivity function of the asymmetric CLP sequence in Fig.\textit{\ref{fig-5}} (b). The atoms always stay in the same state $|a\rangle$ (solid lines) or $|b\rangle$ (dashed lines) during the time intervals of [$t_{1}:t_{2}$] and [$t_{3}:t_{4}$]. The atom interferometer has a high immunity to the phase noise in the above time intervals. We discuss the phase in the time intervals of [$t_{0}:t_{1}$] and [$t_{2}:t_{3}$]. Due to the fact that the Rabi oscillation induced $\pi$-phase flip causes the opposite phase shifts between [$t_{0}:t_{1}$] and [$t_{2}:t_{3}$] in the symmetric CLP sequence, the phase noise ($<10$ kHz) is cancelled. However, the phase noise is accumulated in the asymmetric CLP sequence. When the polarizations of Raman lasers are half-rotated at $t_{2}$ \cite{Explain3}, the sign induced $\pi$-phase flip is introduced, and the sensitivity function is changed from Fig.\textit{\ref{fig-5}} (d) to Fig.\textit{\ref{fig-5}} (c). Thus, the phase noise ($<10$ kHz) in the asymmetric CLP sequence is also cancelled because there are the opposite phase shifts between [$t_{0}:t_{1}$] and [$t_{2}:t_{3}$]. The LMT beam mirrors of the symmetric and asymmetric CLP sequences can be built, and the phase noise ($<10$ kHz) can also be cancelled by the same ways. At the same time, the atoms stay in the same state during the atomic free evolution (i.e., solid lines from $t_{3}$ to $t_{4}$). Thus, the slow varying phase noise both in and between the LMT beam splitters and mirrors can be simultaneously cancelled. The atom interferometer may be not limited by the classical behavior, such as the phase shift of Raman lasers caused by the temperature drift, which is important for improving the sensitivity and especially for the stability of atom interferometer.

In conclusion, the Rabi oscillation induced phase flip is observed for the first time. The $\pi$-phase flip caused by the negative sign of transition amplitude is confirmed. The $\pi$-phase flips observed by modulating the polarization and duration of Raman lasers in this work and by rotating the magnetic field in Refs.\cite{Usami2007a,Takahashi2009a}, are well explained by a universal theoretical model. They arise from the positive or negative sign of $\sin(\Omega_{2}\tau_{2})$ term in Eq.(\ref{eq2}), which are visually depicted on the Bloch sphere. The CLP sequences based LMT beam splitters and mirrors are proposed for building the low-phase-noise large-area atom interferometer. Importantly, the slow varying phase noise both in and between the LMT beam splitters and mirrors can be simultaneously cancelled. Furthermore, fringes can be obtained by varying the intensity or duration of Raman lasers. The CLP sequences can also be depicted on the Bolch sphere, and they may be optimized by controlling the polar angle and azimuthal angle on the Bloch sphere. The influence of atomic velocity distribution can be decreased in the ultracold atom interferometer. This work has the potential application in atom interferometers, and it is important for developing the ultra-sensitivity atom interferometer.

We acknowledge the financial support from the National Natural Science Foundation of China under Grant Nos. 11004227 and 91536221, and funds of Youth Innovation Promotion Association of Chinese Academy of Sciences. We also thank X. W. Guan and J. Luo from WIPM, L. Deng from NIST for helpful suggestions and comments.

\bibliography{basename of .bib file}

\begin{thebibliography}{99}
\bibitem {Note}i.e., the atoms are prepared in a superposition state by the first Raman pulse ($\Omega_{1} \tau_{1}=\pi/2$), and they are recombined by the second pulse ($\Omega_{2} \tau_{2} \not= \pi/2$). $\Omega_{j} \tau_{j}=\pm\pi/2$ are defined as $\pm\pi/2$ pulses for the $j$th interactions.
\bibitem {Explain1}The angle of polarization rotation is 2 times as the angle of polarization axis of the wave plate rotated, $\beta_{2}=2\beta_{2}^{'}$.
\bibitem {Explain2}Each pulse is composed of a pair Raman lasers, and the phase difference of two lasers is zero when their polarizations are simultaneously rotated.
\bibitem {Explain3}i.e., the polarizations of Raman lasers can be half-rotated by a fast electro-optical modulator (several nanoseconds).
\end{thebibliography}

\end{document}